\documentclass[]{llncs}
\usepackage[dvips]{graphicx}

\title{An Efficient Algorithm for Enumerating Chordless Cycles and Chordless Paths}
\author{Takeaki Uno\inst{1} and Hiroko Satoh\inst{1}}
\institute{National Institute of Informatics, 2-1-2 Hitotsubashi, Chiyoda-ku, Tokyo 101-8430, JAPAN, 
{\tt e-mail: \{uno,hsatoh\}@nii.ac.jp}}

\begin{document}
\maketitle

\begin{abstract}
A chordless cycle (induced cycle) $C$ of a graph is a cycle without any chord,
 meaning that there is no edge outside the cycle connecting two vertices
 of the cycle.
A chordless path is defined similarly.
In this paper, we consider the problems of enumerating chordless cycles/paths
 of a given graph $G=(V,E),$ and propose algorithms taking $O(|E|)$ time 
 for each chordless cycle/path.
In the existing studies, the problems had not been deeply studied in 
 the theoretical computer science area, and no output polynomial
 time algorithm has been proposed.
Our experiments showed that the computation time of our algorithms
 is constant per chordless cycle/path for non-dense random graphs and
 real-world graphs.
They also show that the number of chordless cycles is much smaller
 than the number of cycles.
We applied the algorithm to prediction of NMR (Nuclear Magnetic Resonance)
 spectra, and increased the accuracy of the prediction.
\end{abstract}

%%%%%%%%%%%%%%%%%%%%%%%%%%%%%%%%%%%%%%%%%%%%%%%%%%%%%%%%%%%%%%%%%%%%%%
\section{Introduction}

Enumeration is a fundamental problem in computer science, and many
 algorithms have been proposed for many problems, such as cycles, paths,
 trees and cliques\cite{Ep90,KpRm00,MkUn04,RdTj75,tomita,Un01}.
However, their application to real world problems has not been researched
 very much, due to the handling needed for the huge amount of and the 
 high computational cost.
However, this situation is now changing, thanks to the rapid increase
 in computational power, and the emergence of data centric science.
For example, the enumeration of all substructures frequently appearing in
 a database, i.e., frequent pattern mining, has been intensively studied.
This method is adopted for capturing the properties of databases,
 or for discovering new interesting knowledge in databases.
Enumeration is necessary for such tasks because the objectives cannot
 be expressed well in mathematical terms.
The use of good models helps reduce the amount of output, and 
 the use of efficient algorithms enables huge databases to be more easily
  handled\cite{InWsMt03,UNOT,seq}.
More specifically, introducing a threshold value for the frequency, which
 enables controlling the number of solutions.
In such areas, minimal/maximal solutions are also enumerated to
 reduce the number of solutions.
For example, enumerating all cliques is usually not practical while 
 enumerating all maximal cliques, i.e. cliques included in no other
  cliques, is often practical\cite{tomita,MkUn04}.
In real-world sparse graphs, the number of maximal cliques is not exponential,
 so, even in large-scale graphs, the maximal cliques can often be enumerated
  in a practically short time by a stand alone PC even for graphs with millions of vertices.
However, the enumeration of maximum cliques, that have the maximum number
 of vertices among all cliques, is often not acceptable in practice, since
 the purpose of enumeration is to find all locally dense structures,
 and finding only maximum cliques will lose relatively small dense
 structures, thus it does not cover whole the data.

Paths and cycles are two of the most fundamental graph structures.
They appear in many problems in computer science and are used for
 solving problems, such as optimizations (e.g. flow problems) and
 information retrieval (e.g. connectivity and movement of objects).
Paths and cycles themselves are also used to model other objects.
For example, in chemistry, the size and the fusing pattern of cycles in
 chemical graphs, representing chemical compounds, are considered to be
 essential structural attributes affecting on several important properties
 of chemical compounds, such as spectroscopic output, physical property,
 chemical reactivity, and biological activity. 

For a path/cycle $P$, an edge connecting two vertices of $P$ but
 not included in $P$ is called a chord.
A path/cycle without a chord is called a chordless path/cycle.
Since a chordless cycle includes no other cycle as a vertex set,
 it is considered minimal.
Thus, chordless cycles can be used to represent cyclic structures.
For instance, the size and fusing pattern of chordless cycles in chemical
 graphs as well as other properties of chemical structures are taken into
 account when selecting data for prediction of nuclear magnetic resonance
 (NMR) chemical shift values\cite{St}.
Most chemical compounds contain cycles.
In chemistry, the term `ring' is used instead of `cycle', for example
 a cycle consisting of 5 vertices is called 5-membered ring.
Since the character of ring structures of chemical compounds
 is assumed to be important to study the nature of the structure-property
 relationships, the ring perception is one of classical questions
 \cite{chem-Bl,chem-Dw2,chem-Dw,chem-HsJfGf96} in the context of chemical
 informatics, so called chemoinformatics.
Several kinds of ring structures, such as all rings and the smallest set of
 smallest ring (SSSR) are usually included in a basic dataset of chemical
 information.
NMR chemical shift prediction is a successful case where the information
 about chordless cycles is employed to improve the accuracy of the prediction.
The path/cycle enumeration is supposed to be useful also for analysis of
 network systems such as Web and social networks.

In this paper, we consider the problem of enumerating all chordless
 paths (resp., cycles) of the given graph.
While optimization problems for paths and cycles have been studied well,
 their enumeration problems have not.
This is because there are huge numbers of paths and cycles 
 even in small graphs.
However, we can reduce the numbers so that the problem becomes tractable
 by introducing the concept of chordless.
The first path/cycle enumeration algorithm was proposed by Read and Tarjan 
 in 1975\cite{RdTj75}.
Their algorithm takes as input a graph $G=(V,E)$ and enumerates all cycles,
 or all paths connecting given vertices $s$ and $t$, in $O(|V|+|E|)$ time
 for each.
The total computation time is $O((|V|+|E|)N)$ where $N$ is the 
 number of output cycles/paths.
Ferreira et al. \cite{FGMPRS} recently proposed a faster algorithm, 
 that takes time linear in the output size, that is the sum of the
 lengths of the paths.

The chordless version was considered by Wild\cite{Wl08}.
An algorithm based on the principle of exclusion is proposed, but
 the computational efficiency was not considered deeply.
In this paper, we propose algorithms for enumerating chordless cycles
 and chordless paths connecting two vertices $s$ and $t$ (reported in
 2003\cite{Un03}).
Note that chordless cycles can be enumerated by chordless path enumeration.
The running time of the algorithm is $O(|V|+|E|)$ for each, the same as
 the Read and Tarjan algorithm.

We experimentally evaluated the practical performance of the algorithms
 for random graphs and real-world graphs.
The results showed that its practical computation time is much smaller
 than $O(|V|+|E|)$, meaning that the algorithms can be used for large-scale
 graphs with non-huge amount of solutions.
The results also showed that the number of chordless cycles is drastically
 small compared to the number of usual cycles.

\begin{figure}[t]
  \begin{center}
  \includegraphics[scale=0.31]{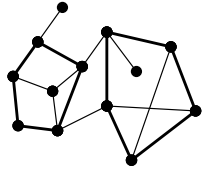}
  \end{center}
  \caption{Left bold cycle is a chordless cycle, right bold cycle
 has two chords.}\label{chordless}
\end{figure}

\vspace{-2mm}
%%%%%%%%%%%%%%%%%%%%%%%%%%%%%%%%%%%%%%%%%%%%%%%%%%%%%%%%%%%%%%%%%%%%%%%
\section{Preliminaries}
\vspace{-2mm}

A {\em graph} is a combination of a vertex set and an edge set such that
 each edge is a pair of vertices.
A graph $G$ with vertex set $V$ and edge set $E$ is denoted by $G=(V,E)$.
An edge $e$ of pair $v$ and $u$ is denoted by $\{u,v\}$.
We say that the edge {\em connects} $u$ and $v$, $e$ is {\em incident}
 to $u$ and $v$, and $v$ and $u$ are {\em adjacent} to each other, and 
 call $u$ and $v$ {\em end vertices} of $e$.
An edge with end vertices that are the same vertex is called a {\em self-loop}.
Two edges having the same end vertices $u$ and $v$ are called
 {\em multi-edges}.
We deal only with graphs with neither a self-loop nor a multi-edge.
This restriction does not lose the generality of the problem formulation.

A path is a graph of vertices and edges composing a sequence
 $v_1,\{ v_1,v_2\},v_2,$\\
 $\{ v_2,v_3\},\ldots,\{ v_{k-1}, v_k\},v_k$
  satisfying $v_i \ne v_j$ and $i\ne j$.
The $v_1$ and $v_k$ are called the {\em end vertices} of the path.
If the end vertices of $P$ are $s$ and $t$, the path is called an
 $s$-$t$ path.
When $v_1= v_k$ holds, a path is called a cycle.
Here we represent paths and cycles by vertex sequences,
 such as $(v_1,\ldots,v_k)$.
An edge connecting two vertices of a path/cycle $P$ and not included in
 $P$ is called a {\em chord} of $P$.
A path/cycle $P$ such that the graph includes no chord of $P$ is called
 {\em chordless}.
Figure \ref{chordless} shows examples.
In a set system composed of the vertex sets of cycles (resp.,
 $s$-$t$ paths), the vertex set of a chordless cycle (resp., $s$-$t$ path)
 is a minimal element.

For a graph $G$ and a vertex subset $S$ of $G$, $G\setminus S$ denotes 
 the graph obtained from $G$ by removing all vertices of $S$ and all
 edges incident to some vertices in $S$.
For a vertex $v$, $N(v)$ denotes the {\em neighbor} of $v$, that is, 
 the set of vertices adjacent to $v$.
For a vertex set $S$ and a vertex $v$, $S\setminus v$ and $S\cup v$ 
 denote $S\setminus \{ v\}$ and  $S\cup \{ v\}$, respectively.
For a path $P$ and its end vertex $v$, $P\setminus v$ denotes the path 
 obtained by removing $v$ from $P$.

%%%%%%%%%%%%%%%
\begin{property}\label{st-exist}
There is a chordless $s$-$t$ path if and only if there is an $s$-$t$ path.
\end{property}

\proof
A chordless $s$-$t$ path is an $s$-$t$ path, thus only if part is true.
If an $s$-$t$ path exists, a shortest path from $s$ to $t$ is a 
 chordless $s$-$t$ path, and thus it always exists.
\qed
%%%%%%%%%%%%%%%%%

%%%%%%%%%%%%%%%
\begin{property}\label{exist}
A vertex $v$ is included in a cycle if and only if $v$ is included in 
 a chordless cycle.
\end{property}

\proof
If $v$ is not included in any cycle, it obviously is not included in 
 any chordless cycle.
Hence, we investigate the case in which $v$ is included in a cycle $C$.
If $C$ is chordless, we are done.
If $C$ has a chord, the addition of the chord splits $C$ into
 two smaller cycles, and $v$ is always included in one of them.
We then consider the cycle as $C$.
The cycle with three vertices can not have a chord, thus 
 we always meet a chordless cycle including $v$.
\qed
%%%%%%%%%%%%%%%%%

For a recursive algorithm, an iteration means the computation 
 from the beginning of a recursive call to its end, excluding any 
 computation done in recursive calls generated in the iteration.
If an iteration $I$ recursively calls an iteration $I'$, 
 $I'$ is called a {\em child} of $I$, 
 and $I$ is called the {\em parent} of $I'$.

\vspace{-2mm}
%%%%%%%%%%%%%%%%%%%%%%%%%%%%%%%%%%%%%%%%%%%%%%%%%%%%%%%%%%%%%%%%%%%%%%
\section{Algorithm for Chordless $s$-$t$ Path Enumeration}
\vspace{-2mm}

Our enumeration problem is formulated as follows.\\

\noindent
{\bf Chordless $s$-$t$ path enumeration problem}\\
For a given graph $G=(V,E)$ and two vertices $s$ and $t$, enumerate 
all chordless $s$-$t$ paths included in $G$.\\

We first observe that chordless cycle enumeration is done with 
 chordless $s$-$t$ path enumeration by repeating steps;
 (1) for a vertex $s$, enumerate chordless $s$-$t$ paths in $G\setminus \{ s,t\}$ for each vertex $t$ adjacent to $s$, and
 (2) remove $s$ from the graph.
Here $G\setminus \{ s,t\}$ is the graph obtained from $G$ by removing the
 edge $\{ s,t\}$.
This implies that we only have to consider chordless $s$-$t$ path enumeration.

\begin{lemma}\label{st}
For a vertex $v\in N(s)$, 
 $P$ is a chordless $s$-$t$ path including $v$ if and only if 
 $P\setminus s$ is a chordless $v$-$t$ path of the graph
 $G\setminus (N(s) \setminus v)$.
\end{lemma}

\proof
If $P\setminus s$ is a chordless $v$-$t$ path in
 $G\setminus (N(s)\setminus v)$, $P$ is an $s$-$t$ path all whose chords
 are incident to $s$.
Since $P$ has no vertex in $N(s) \setminus v$, no vertex of $P$
 other than $v$ is adjacent to $s$. 
Thus, $P$ has no chord incident to $s$, and is chordless.

If $P$ is a chordless $s$-$t$ path including $v$, no vertex 
 $u\in N(s)\setminus v$ is included in $P$, since 
 the edge $\{ s,u\}$ would be a chord if was included.
Thus, $P\setminus s$ is a chordless $v$-$t$ path in
 $G\setminus (N(s) \setminus v)$.
\qed

\begin{lemma}
The set of chordless $s$-$t$ paths of $G$ is partitioned into disjoint
 sets of chordless $s$-$t$ paths in the graphs
 $G\setminus (N(s) \setminus v)$ for each $v$.
\end{lemma}

\proof
Suppose that $P$ is a chordless $s$-$t$ path in $G$.
Then, from lemma \ref{st}, $P$ includes exactly one vertex among $N(s)$.
If $P$ includes $v\in N(s)$, $P\setminus s$ is a chordless $v$-$t$ path
 in $G\setminus (N(s) \setminus v)$, thus $P$ is a chordless $s$-$t$ path
 in $G\setminus (N(s) \setminus v)$.
Since $P$ is not an $s$-$t$ path in $G\setminus (N(s) \setminus u)$ 
 for any $u\in N(s) \setminus v$, the statement holds.
\qed

From the lemma, we obtain the following algorithm.
The $Q$ is the sequence of vertices attached to the paths in the ancestor
 iterations, and set to be empty at the start of the algorithm.

\begin{tabbing}
{\bf Enum\_Chordless\_Path} ($G=(V,E), s, t, Q$)\\
1. {\bf if} edge $\{ s, t\}$ exists in $E$ {\bf then} output $Q \cup t$;
 {\bf return}\\
2. {\bf for each} $v\in N(s)$ s.t. a $v$-$t$ path exists in
 $G\setminus (N(s) \setminus v)$ {\bf do}\\
3. \ \  \ \ call {\bf Enum\_Chordless\_Path} ($(G\setminus (N(s)\setminus
  v))\setminus s, v, t, Q\cup v$)\\
4. {\bf end for}
\end{tabbing}

\begin{figure}[t]
\vspace{-5mm}
  \begin{center}
  \includegraphics[scale=0.31]{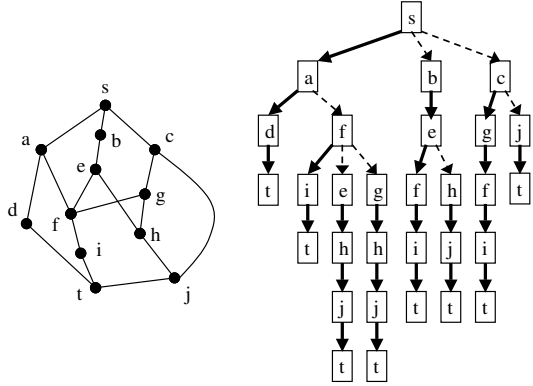}
  \end{center}
  \caption{Tree on the right represents recursive structure of
  $s$-$t$ path enumeration in the graph on left; bold lines
   correspond to recursive calls in step 2, and dotted lines correspond 
   to those in step 6.}\label{p2}
\end{figure}

When a recursive call is generated in an iteration of the algorithm, 
 $G\setminus (N(s)\setminus v)$ is generated from $G$ by removing vertices
 and edges. 
The removed vertices and edges are kept in memory so that $G$ can be 
 reconstructed from the graph.
A removed edge or vertex is not removed again in the descendants
 of the iteration.
Thus, the accumulated memory usage for these removed vertices and edges is
 $O(|V|+|E|)$, and the space complexity of the algorithm is $O(|V|+|E|)$.
 
In step 2, all vertices $v\in N(s)$ such that a $v$-$t$
 path exists in $G\setminus (N(s) \setminus v)$ must be listed. 
If and only if the condition in step 2 holds, there is a vertex $u\in N(v)$
 such that a $u$-$t$ path exists in $(G\setminus N(s))\setminus s$.
Thus, those vertices can be listed by computing the connected
 component including $t$ in $G\setminus N(s)\setminus s$ and 
 checking the condition in step 2 for all $u\in N(v)$ for all $v\in N(s)$.
This can be done in $O(|V|+|E|)$ time.
The construction of $(G\setminus (N(s)\setminus v))\setminus s$
 is done in $O(|N(v)|)$ time by constructing it from
 $(G\setminus N(s))\setminus s$.
Therefore, the time complexity of an iteration is $O(|V|+|E|)$.

Let us consider the recursion tree of the algorithm which is a tree
 representing the recursive structure of the algorithm.
The vertex of the recursion tree corresponds to an iteration, and 
 each iteration and its parent are connected by an edge.
The leaves correspond to the iterations generating no recursive calls, 
 and the algorithm outputs a solution on each leaf.
Because of the condition given placed on vertices in step 2, there is
 always at least one $s$-$t$ path in the given graph.
This implies that at least one recursive call occurs when step 2 is executed.
Hence, the algorithm outputs a solution at every leaf of the recursion tree.
The depth of the recursion tree is $O(|V|)$ since at least one vertex is 
 removed from the graph to generate a recursive call.
We can conclude from these observations that the time complexity of the 
 algorithm is $O(N|V|(|V|+|E|))$ where $N$ is the number of 
 chordless $s$-$t$ paths in $G$.
Next, we discuss the reduction of the time complexity to $O(N(|V|+|E|))$.

We first rewrite the above algorithm as follows.
We denote the vertex next to $v$ in path $P$ by $nxt(v)$.
Note that although we introduce several variables, the algorithms
 are equivalent.

\begin{tabbing}
{\bf Enum\_Chordless\_Path2} ($G=(V,E), s, t, Q$)\\
1. {\bf if} $s$ is adjacent to $t$ {\bf then output} $Q\cup t$ ; {\bf return}\\
2. $P := $ a chordless $s$-$t$ path in $G$\\
3. {\bf call} {\bf Enum\_Chordless\_Path2} ($G\setminus (N(s)\setminus nxt(s)),
 nxt(s), t, Q\cup nxt(s)$ )\\
4. {\bf for each} $v\in N(s), v\ne nxt(s)$ {\bf do}\\
5. \ \ {\bf if} there is a $v$-$t$ path in
      $G\setminus (N(s)\setminus v)$ {\bf then}\\
6. \ \ \ \ {\bf call} {\bf Enum\_Chordless\_Path2}
    ( $G\setminus (N(s)\setminus v), v, t, Q\cup v$ )\\
7. {\bf end for}
\end{tabbing}

We further rewrite the algorithm as follows.
We compute the chordless $s$-$t$ path $P$ computed in step 2 of the
 above algorithm, before the start of the iteration, i.e., in its parent,
 and give it as a parameter to the recursive call.
 
\begin{tabbing}
{\bf Enum\_Chordless\_Path3} ($G=(V,E), s, t, Q, P$)\\
1. {\bf if} $s$ is adjacent to $t$ {\bf then output} $Q\cup t$ ; {\bf return}\\
2. {\bf call} {\bf Enum\_Chordless\_Path3} ($G\setminus (N(s)\setminus nxt(s)),
 nxt(s), t, Q\cup nxt(s), P\setminus s$)\\
3. {\bf for each} $v\in N(s), v\ne nxt(s)$ {\bf do}\\
4. \ \ {\bf if} there is an $v$-$t$ path in 
      $G\setminus (N(s)\setminus v)$ {\bf then}\\
5. \ \ \ \ $P :=$ a chordless $v$-$t$ path in $G\setminus (N(s)\setminus v)$ (found by a breadth first search)\\
6. \ \ \ \ {\bf call} {\bf Enum\_Chordless\_Path3}
    ($G\setminus (N(s)\setminus v), v, t, Q\cup v, P$)\\
7. \ \ {\bf end if}\\
8. {\bf end for}
\end{tabbing}
 
Figure \ref{p2} illustrates an example of the recursive structure
 of this algorithm.
The tail of an arrow is a parent and the head is its child.
We call the child generated in step 2 {\em first child}, and the arrow 
 pointing at the first child is drawn with a bold line.
We can make a path by following the bold-arrows, and we call a maximal such
 path a {\em straight path}.
Since the bottom of a straight path is a leaf, the number of straight paths
 is bounded by the number of chordless paths.
Since the head of a non-bold arrow always points an end of a straight path,
 the number of non-bold arrows, that correspond to the recursive calls done
 in step 6, is bounded by the number of straight paths.

From these observations, we infer the following points regarding
 time complexity.

\vspace{-1mm}
\begin{itemize}
\item An iteration takes $O(|V|+|E|)$ time when a chordless path is output.
 This computation time is $O(|V|+|E|)$ per chordless path. 

\item Steps 1 and 2 take $O(NN(s))$ time where $NN(s)$ is the 
number of edges adjacent to vertices in $N(s)$.
This time is spent checking the adjacency of $s$ and $t$ and 
 constructing $G\setminus (N(s)\setminus v)$ for all $v\in N(s)$.
This comes from that $G\setminus (N(s)\setminus v)$ can be constructed
 from $G\setminus N(s)$ by adding edges adjacent to $v$ in $O(|N(v)|)$ time.

\item The number of executions of the for loop in step 3 is 
 bounded by $|N(s)|$.
Their sum over all iterations in a straight path does not exceed
 the number of edges.

\item Steps 5 and 6 take $O(|V|+|E|)$ time to find a chordless $v$-$t$ path, 
 and to construct $G\setminus (N(s)\setminus v)$. 
Since the recursive call in step 6 corresponds to a straight path, 
 this computation time is $O(|V|+|E|)$ per chordless path.

\item The execution time for step 4 is $O(|V|+|E|)$.
\end{itemize}
\vspace{-1mm}

We see from the above that the bottle neck in terms of time complexity is
 step 4.
The other parts of the algorithm takes $O(|V|+|E|)$ time
 per chordless $s$-$t$ path.
We speed up step 4 by using the following property.

\begin{property}
$G\setminus \{v \}$ includes a $v$-$t$ path for $v\in N(s)$ if and only if
 there is a vertex $u\in N(v)\setminus N(s)$ such that $G\setminus N(s)$
 includes a $u$-$t$ path.
\qed
\end{property}

In each iteration we put mark on vertices $u$ such that there is a $u$-$t$
 path in $G\setminus N(s)$.
Step 4 is then done in $O(|N(v)|)$ time by looking at the marks on 
 the vertices in $N(v)$.
The marks can be put in short time, by updating the marks put in the
 first child.
The condition of step 4 is checked by finding all vertices in
 $G\setminus N(s)$ from which going to $t$ is possible.
This also takes $O(|V|+|E|)$ time, but the time is reduced by re-using
 the results of the computation done for the first child.
In the first child, marks are put according to the reachability to $t$
 in $G\setminus (N(s)\cup N(nxt(s)))$.
To put marks for $G\setminus N(s)$, we find all vertices $u$ such that 
 any $u$-$t$ path in $G\setminus N(s)$ includes a vertex of
 $N(nxt(s)) \setminus N(s)$.
This is done by using a graph search starting from the vertices of
 $N(nxt(s)) \setminus N(s)$ that are adjacent to a marked vertex, and
 visiting only unmarked vertices.
The time taken is linear in the number of edges adjacent to newly
 marked vertices.

Consider the computation time with respect to step 4, for the iterations
 in a straight path.
In these operations, a vertex (resp., an edge) gets a mark at most once, 
 i.e., it never gets a mark twice.
Thus, the total computation time for this computation is
 linear in the sum of the degrees of marked vertices and vertices in
  $N(nxt(s))$, and is bounded by $O(|V|+|E|)$.
The computation time for step 4 is thus reduced to $O(|V|+|E|)$ per
 chordless $s$-$t$ path.
When a recursive call for a non-first child is made, all marks are deleted.
We then perform a graph search starting from $t$ to put the marks.
Both steps take $O(|V|+|E|)$ time.
Since this computation is done only when generating non-first child, the 
 total number of occurrences of this computation is bounded by the number of 
 maximal paths, i.e., the number of chordless paths.
Thus, this computation takes $O(|V|+|E|)$ time for each chordless path.
The algorithm is written as follows.

\begin{tabbing}
{\bf Enum\_Chordless\_Path4} ($G=(V,E), s, t, Q, P$)\\
1. {\bf if} $s$ is adjacent to $t$ {\bf then output} $Q\cup t$; {\bf go to} 11\\
2. {\bf call} {\bf Enum\_Chordless\_Path4} ($G\setminus (N(s)\setminus nxt(s)),
  nxt(s), t, Q\cup nxt(s), P\setminus s$ )\\
3. put mark by graph search on $G\setminus N(s)$ from vertices in $N(nxt(s))$\\
4. {\bf for each} $v\in N(s), v\ne nxt(s)$ {\bf do}\\
4. \ \ {\bf if} a vertex adjacent to $v$ is marked {\bf then}\\
5. \ \ \ \ delete marks from all vertices in $G$\\
6. \ \ \ \ $P :=$ a chordless $v$-$t$ path in $G\setminus (N(s)\setminus v)$\\
7. \ \ \ \ {\bf call} {\bf Enum\_Chordless\_Path4}
    ($G\setminus (N(s)\setminus v), v, t, Q\cup v, P$)\\
8. \ \ \ \ recover the marks deleted in step 5, by graph search starting from $t$ on $G\setminus N(s)$\\
9. \ \ {\bf end if}\\
10. {\bf end for}
\end{tabbing}

\vspace{-2mm}
\begin{theorem}
The chordless $s$-$t$ paths in a given graph $G=(V,E)$ can be enumerated in 
 $O(|V|+|E|)$ time per chordless path, in particular, polynomial
  time delay.
\end{theorem}

\proof
We can see the correctness in the above.
The time complexity of an iteration is $O(|V|+|E|)$, and each iteration
 outputs an $s$-$t$-path.
Moreover, the height of the recursion tree is at most $|V|$, thus 
 the time between two consecutive output paths is bounded by 
 $O(|V|+|E|) + O(|V|) = O(|V|+|E|)$.
This concludes the theorem.
\qed

\begin{theorem}
The chordless cycles in a given graph $G=(V,E)$ can be enumerated in 
 $O(|V|+|E|)$ time per chordless cycle, in particular, polynomial time delay.
 \qed
\end{theorem}

\begin{table}[t]
\begin{center}
\caption{Computational time (in seconds) for randomly generated graphs}\label{time}
\vspace{-2mm}
\small
\begin{tabular}{|r|r|r|r|r|r|r|r|r|r|r|}
 \hline
\small
 edge density  & 10\% & 20\% & 30\% & 40\% & 50\% & 60\% & 70\% & 80\% & 90\% \\ \hline
no. of vertices 50 & 0.18 & 0.12 & 0.098 & 0.089 & 0.082 & 0.08 & 0.085 & 0.1 & 0.11 \\
 75 & 0.17 & 0.12 & 0.099 & 0.088 & 0.079 & 0.074 & 0.077 & 0.088 & 0.1 \\
 100 & 0.17 & 0.12 & 0.099 & 0.09 & 0.083 & 0.081 & 0.089 & 0.095 & 0.12 \\
 150 & 0.2 & 0.12 & 0.099 & 0.098 & 0.077 & 0.075 & 0.083 & 0.103 & 0.14 \\
 200 & 0.18 & 0.12 & 0.1 & 0.088 & 0.081 & 0.078 & 0.085 & 0.11 & 0.17 \\
 300 & 0.19 & 0.12 & 0.1 & 0.087 & 0.082 & 0.083 & 0.091 & 0.12 & 0.21 \\
 400 & 0.17 & 0.12 & 0.1 & 0.089 & 0.08 & 0.086 & 0.1 & 0.15 & 0.26 \\
 600 & 0.18 & 0.11 & 0.12 & 0.12 & 0.11 & 0.1 & 0.13 & 0.23 & 0.42 \\
 800 & 0.2 & 0.12 & 0.14 & 0.13 & 0.11 & 0.11 & 0.13 & 0.26 & 0.54 \\
 1200 & 0.23 & 0.17 & 0.17 & 0.13 & 0.12 & 0.12 & 0.15 & 0.28 & 1 \\
 1600 & 0.24 & 0.19 & 0.14 & 0.13 & 0.13 & 0.14 & 0.21 & 0.29 & 1.3 \\
 2400 & 0.25 & 0.19 & 0.17 & 0.15 & 0.16 & 0.16 & 0.19 & 0.44 & 1.4 \\
 3200 & 0.29 & 0.23 & 0.2 & 0.19 & 0.18 & 0.2 & 0.25 & 0.61 & 1.79 \\
 4800 & 0.28 & 0.28 & 0.27 &  &  &  &  &  & \\ \hline
\end{tabular}
\end{center}
\end{table}

%%%%%%%%%%%%%%%%%%%%%%%%%%%%%%%%%%%%%%%%%%%%%%%%%%%%%%%%%%%%%%%%%%%%%%
\section{Computational Experiments}

The practical efficiency of the proposed algorithms is evaluated 
 by computational experiments.
The results were compared with those of the cycle enumeration algorithm
 proposed in \cite{RdTj75}.
The difference between the number of cycles and of chordless cycles was
 also compared.
The program was coded in C, and compiled using gcc.
The experiments were done on a PC with a Core i7 3GHz CPU.
The code is available at the author's web site
 (http://research.nii.ac.jp/\~uno/codes.html).
We did not use multiple cores, and the memory usage was less than 4MB.
The instance graphs were random graphs and the real-world graphs taken
 from the UCI machine learning repository\cite{UCI}.
All the test instances shown here are downloadable from the author's web
 site, except for those from UCI repository.
Tables \ref{time} to \ref{realworld} summarize the computation time,
 number of cycles, and number of chordless cycles for each instance,
 and clarify the effectiveness of the chordless cycle model and our algorithm.
 
The computation time results for randomly generated graphs are shown in
 Table \ref{time}.
The edge density means the probability of being connected by an edge
 for any two vertices.
Execution of an enumeration algorithm involves many iterations with
 different input graphs, thus we thought that there are sufficiently
 many samples even in one execution of the algorithm.
Therefore, we generated just one instance for each parameter.
Each cell represents the computation time needed for 10,000 cycles or
 chordless cycles.
When the computation time was too long so that the number of output
 cycles exceeded one million, we stopped the computation.

When the edge density was close to 100\%, almost all the chordless
 cycles were triangles.
In this case, intuitively, the algorithm spent $O(|V||E|) = O(|V|^3)$ 
 time to find $O(|V|^2)$ chordless cycles.
In contrast, it took almost constant time for each
 chordless cycle in sparse graphs. 
This is because the graph was reduced by repeated recursive calls, 
 and at the bottom levels, the graph sizes were usually constant.

\tabcolsep=0.7mm

\begin{table}[t]
\begin{center}
\caption{Number of chordless cycles (upper) and of cycles (lower)}\label{num}
\end{center}
\vspace{-5mm}
\small
\hspace*{-5mm}
\begin{tabular}{|r|r|r|r|r|r|r|r|r|r|r|}
 \hline
 edge density & 10\% & 20\%  & 30\% & 40\% & 50\% & 60\% & 70\% & 80\% & 90\%
 & 100\%\\ \hline
%10 & 1 & 3 & 14 & 20 & 41 & 45 & 63 & 81 & 120 & \\
%   & 0 & 1 & 4 & 116 & 352 & 2302 & 3697 & 24016 & 108906 & \\ \hline
%15 & 0 & 10 & 34 & 116 & 165 & 193 & 247 & 297 & 350 & 455 \\
%   & 0 & 36 & 1470 & 613255 & 6620995 & 55525881 & - & - & - & - \\ \hline
%20 & 1 & 78 & 298 & 523 & 637 & 752 & 771 & 846 & 908 & 1140 \\
% & 1 & 56348 & 9114083 & - & - & - & - & - & - & - \\ \hline
%25 & 8 & 1218 & 2049 & 2387 & 2099 & 1891 & 1775 & 1854 & 1928 & 2300 \\ \hline
%50 & 64383 & 395233 & 267435 & 146939 & 82607 & 49800 & 34137 & 23788 & 18873 & 19600 \\ \hline
%75 & 119379652 & 69357503 & 14504338 & 3679823 & 1158210 & 465368 & 221990 & 119719 & 73810 & 67525\\ \hline
%100 & - & - & - & 40436716 & 8269935 & 2395333 & 877466 & 393549 & 199133 & 161700 \\ \hline
%150 & - & - & - & - & 149022507 & 27483087 & 6641331 & 2167686 & 854209 & 551300 \\ \hline
%200 & - & - & - & - & - & 167408239 & 30334867 & 7755298 & 2466694 & 1313400 \\ \hline
%300 & - & - & - & - & - & - & - & 51043836 & 11457502 & 4455100 \\ \hline
%400 & - & - & - & - & - & - & - & - & 35154412 & 10586800 \\ \hline
%
no. of vertices 10 & 1 & 3 & 14 & 20 & 41 & 45 & 63 & 81 & 120 & \\
   & 0 & 1 & 4 & 116 & 352 & 2302 & 3697 & 24k & 108k & \\ \hline
 15 & 0 & 10 & 34 & 116 & 165 & 193 & 247 & 297 & 350 & 455\\
   & 0 & 36 & 1470 & 613k & 6620k & 55525k & - & - & - & - \\ \hline
20 & 1 & 78 & 298 & 523 & 637 & 752 & 771 & 846 & 908 & 1140 \\
 & 1 & 56k & 9114k & - & - & - & - & - & - & - \\ \hline
25 & 8 & 1218 & 2049 & 2387 & 2099 & 1891 & 1775 & 1854 & 1928 & 2300 \\ \hline
50 & 64k & 395k & 267k & 146k & 82k & 49k & 34k & 23k & 18k & 19k \\ \hline
75 & 119379k & 69357k & 14504k & 3679k & 1158k & 465k & 221k & 119k & 73810 & 67525\\ \hline
100 & - & - & - & 40436k & 8269k & 2395k & 877k & 393k & 199k & 161k \\ \hline
150 & - & - & - & - & 149022k & 27483k & 6641k & 2167k & 854k & 551k \\ \hline
200 & - & - & - & - & - & 167408k & 30334k & 7755k & 2466k & 1313k \\ \hline
300 & - & - & - & - & - & - & - & 51043k & 11457k & 4455k \\ \hline
400 & - & - & - & - & - & - & - & - & 35154k & 10586k \\ \hline
\end{tabular}
\end{table}

Table \ref{num} shows that the number of chordless cycles
 exponentially increased against with the edge density, but not
 as much as usual cycles.
Table \ref{sparse} shows the experimental results for
 sparse graphs.
The graphs were generated by adding chords randomly to a cycle of $n$
 vertices so that the average degree was four.
These sparse graphs included so many chordless cycles.
The graphs with at most 100 vertices were solved in a practically short
 time, and the computation time for each chordless cycle were almost the same.

\begin{table}[t]
\begin{center}
\caption{Computation time and number of chordless cycles for sparse graphs}
\label{sparse}
\vspace{-5mm}
\begin{tabular}{|r|r|r|r|r|r|r|r|r|r|}
 \hline
graph size (no. of vertices) & 10 & 20 & 30 & 40 & 50 & 60 & 70 & 80 & 90 \\ \hline
no. of chordless cycles & 12 & 90 & 743 & 5371 & 89164 & 853704 & 4194491 & 45634757 & - \\
time for 10,000 chordless cycles & 16.6 & 3.33 & 0.53 & 0.22 & 0.18 & 0.2 & 0.23 & 0.24 & - \\ \hline
\end{tabular}
\end{center}
\end{table}

Table \ref{realworld} shows the number of chordless cycles with limited lengths including a vertex (the first vertex) for the real-world data, taken
 from the UCI repository.
The number of all chordless cycles is shown at the bottom for reference.
The graphs were basically sparse, and globally well connected, and thus
 included a large number of cycles.
Even in such cases, by giving an upper bound of the length, 
Some graphs can be made tractable in such cases by placing an upper bound
 on the length.
These results show the possibility of using chordless cycles with limited 
 lengths for practical data analysis of real-world graphs such as those for
 social networks.

\begin{table}
\begin{center}
\caption{No. of chordless cycles including a vertex (of ID $0$), for
 real-world graphs}
\label{realworld}
%\hspace*{-10mm}
\begin{tabular}{|r|r|r|r|r|r|r|r|r|}
 \hline
& adjnoun & astro-ph & breast &celegen & cond-mat-2005 & cond-mat\_large & dolphins \\ \hline
(no. of vertices) & 114 & 16708 & 7539 & 298 & 40423 & 30561 & 64 \\ 
(no. of edges) & 425 & 121251 & 5848 & 2359 & 175693 & 24334 & 159 \\ \hline
length $<5$ & 8 & 327 & - & 3342 & 393 & 6 & 26 \\
length $<8$  & 251 &  & - & 1738k & - & 6 & 320 \\
length $<16$ & 65350 & - & - & - & - & 6 & 1780 \\
\#chord. cyc. & 66235k & - & - & - & - & - & 6966 \\ \hline
\end{tabular}
%\hspace*{-10mm}
\begin{tabular}{|r|r|r|r|r|r|r|r|r|}
 \hline
 & football& human\_ppi & karate & lesmis & netscience & polblogs & polboopks & power\\ \hline
(no. of vertices) & 117 & 10347 & 36 & 79 & 1591 & 1492 & 107 & 4943 \\ 
(no. of edges) & 616 & 5418 & 78 & 254 & 2742 & 19090 & 441 & 6594 \\ \hline
length $<5$ & 81 & 1838 & 37 & 3 & 1 & 35881 & 21 & 0 \\
length $<8$ & 11869 & - & 38 & 3 & 1 & - & 187 & 4 \\
length $<16$ & 256664k  & - & 38 & 3 & 1 & - & 34742 & 60 \\
\#chord. cyc. & - & - & 103 & 594 & 5760 & - & 2273k & - \\ \hline
\end{tabular}
\end{center}
\end{table}

\vspace{-2mm}
%%%%%%%%%%%%%%%%%%%%%%%%%%%%%%%%%%%%%%%%%%%%%%%%%%%%%%%%%%%%%%%%%%%%%%
\subsection{Application to NMR Prediction}
\vspace{-2mm}

Chordless cycle enumeration has already been implemented as a part of 
 a database system of chemoinformatics\cite{St}, composed of
 structural data of chemical compounds.
In this system, the number of chordless cycles in the chemical graph
 of a chemical compound is considered to be an attribute of the compound.
In response to a query about the chemical structure of a compound, the system
 searches in the database for structures partially similar to the structure
 of query compound, and predict some functions of the query compound.
A chemical graph is usually sparse and is globally a tree or a combination
 of several large cycles.
Small components can be attached to the large cycles.
Thus, the number of chordless cycles is not so huge and is tractable 
 in most cases.

The program code was implemented in the CAST/CNMR system for predicting
 the $^{13}$C-NMR chemical shift\cite{St,StKsUzNk}.
The codes and a more precise description of this system are available at
 http://research.nii.ac.jp/\~hsatoh/subjects/NMR-e.html.
The information obtained about chordless cycles is used to improve
 the accuracy for the predicted values when the ring attributes
 affects the NMR spectrum.
The CAST/CNMR system predicts chemical shifts by using a chemical
 structure-spectrum database, containing mainly natural organic products
 and their related synthetic compounds.
Since most of the compounds include chains of fused rings,
 enumerating all rings for these compounds would greatly increase 
 the output size, with lots of data useless for NMR prediction.
Therefore, the chordless cycle was adopted as a relevant ring attribute 
 for the CAST/CNMR system.
For accurate NMR prediction for carbon atoms, an error within 1.0 ppm
 (parts per million) is generally required.
Use of chordless cycle information reduced error values of -4.1 to 1.6
 ppm for some problematic carbon atoms to less than 1.0 ppm\cite{St}.

\vspace{-2mm}
%%%%%%%%%%%%%%%%%%%%%%%%%%%%%%%%%%%%%%%%%%%%%%%%%%%%%%%%%%%%%%%%%
\section{Conclusion}
\vspace{-2mm}

We proposed an algorithm for enumerating all chordless $s$-$t$ paths, that 
 is applicable to chordless cycle enumeration without increasing
 the time complexity.
By reusing the results of the subroutines, the computation time is reduced
 to $O(|V|+|E|)$ for each chordless path.
The results of the computational experiments showed that the algorithm
 works well for both random graphs and real-world graphs; the computation
 time was $O(|V|)$ in dense graphs, and almost constant for sparse graphs.
The results also showed that the number of chordless cycles
 is small compared to the number of usual cycles.
This algorithm thus paves the way to efficient use of cycle
 enumeration in data mining.

\end{document}